\documentclass[a4paper]{article}

\usepackage{hyperref}
\usepackage{amsmath}
\usepackage{mathrsfs}
\usepackage{amssymb}
\usepackage{mathtools}

\usepackage[top=2.8cm,bottom=3cm,left=3cm,right=3cm]{geometry}
\usepackage{graphicx}
\usepackage{txfonts}
\usepackage{amsfonts}
\usepackage{amssymb}
\usepackage{booktabs}
\usepackage{xcolor}
\usepackage{lineno}
\usepackage{natbib}
\bibpunct{(}{)}{;}{a}{}{,} % to follow the A&A style
\begin{document}

%\nolinenumbers

\thispagestyle{empty} 

\begin{center}
{\bf{Extended Drag-Based Model for better predicting the evolution of Coronal Mass Ejections}}
\end{center}

Mattia Rossi$^{1}$, Sabrina Guastavino$^{1,2}$, Michele Piana$^{1,2}$ and Anna Maria Massone$^{1}$ \\

\hspace{-0.5cm}$^{1}$MIDA, Dipartimento di Matematica Università di Genova, via Dodecaneso 35 16146 Genova, Italy\\
$^{2}$Osservatorio Astrofisico di Torino, Istituto Nazionale di Astrofisica, Strada Osservatorio 20 10025, Pino Torinese, Italy

\thispagestyle{empty} 

\begin{abstract}
The solar wind drag-based model is a widely used framework for predicting the propagation of Coronal Mass Ejections (CMEs) through interplanetary space. This model primarily considers the aerodynamic drag exerted by the solar wind on CMEs. However, factors like magnetic forces, pressure gradients, and the internal dynamics within CMEs justify the need of introducing an additional small-scale acceleration term in the game. Indeed, by accounting for this extra acceleration, the extended drag-based model is shown to offer improved accuracy in describing the evolution of CMEs through the heliosphere and, in turn, in forecasting CME trajectories and arrival times at Earth. This enhancement is crucial for better predicting Space Weather events and mitigating their potential impacts on space-based and terrestrial technologies.
\end{abstract}

\hspace{-0.5cm}Keywords: coronal mass ejections -- interplanetary propagation -- drag-based model -- accelerated dynamics -- spacecraft alignment

\section{Introduction } \label{sec:intro}
Coronal Mass Ejections \cite[CMEs;][]{2011LRSP....8....1C,howard2011coronal,2012LRSP....9....3W,2019SciA....5.7004G} are massive outbursts of magnetized plasma from the solar corona into the interplanetary space. When directed toward Earth, they cause severe geomagnetic disturbances \citep{2016GSL.....3....8G,2016ApJ...820...16J,2023ApJ...952..111T,2024ApJ...971...94G} and can pose a persistent hazard as harmful radiation to space and ground-based facilities, and human health. Therefore, predicting the CMEs' arrival time and impact speed to Earth is essential in the context of the Space Weather forecasting science \citep{2019SpWea..17.1166C}.

One of the most popular and commonly used approaches to predict the transit time of a CME and its speed to Earth is known as the Drag-Based Model (DBM) \citep[e.g.,][]{2002JGRA..107.1019V,2013SoPh..285..295V,2018JSWSC...8A..11N,2018ApJ...854..180D}. This model assumes that the kinematics of the CME is governed by its dynamic interaction with the Parker spiral-shaped interplanetary structures (i.e., high- and low-speed streams) where it propagates, via the magnetohydrodynamic (MHD) equivalent of the aerodynamic drag force. The model, which mathematically reduces to a rather simple equation of motion, thus essentially predicts that the speed of the CME will balance that of the ambient solar wind in which it is expanding. Recent efforts have also been devoted to incorporating the physics of aerodynamic drag into methodologies based on Artificial Intelligence (AI) techniques, paving the way for innovative (hybrid) approaches known as physics-driven AI models \citep{2023ApJ...954..151G}.

Although the DBM has been subject to continuous refinements \citep[in this regard, it is worth mentioning the $3$D COronal Rope Ejection (3DCORE) developed by][]{2018SpWea..16..216M} and is now a well-established approach, the several drawbacks associated with its intrinsic approximations are evident. Indeed, it is clear that the complex dynamical interaction of the CME with its surroundings cannot be properly described solely by the drag force. Other important physical processes are certainly at play in the evolution of CMEs: these include CME rotation, reconfiguration, deformation, deflection, erosion along with any other magnetic reconnection-driven processes \citep[see the review by][for a rather comprehensive dissertation on this topic]{2017SSRv..212.1159M}, resulting in additional accelerations beyond that predicted by the trivial DBM acting on the CME as it travels through the heliosphere.

\citet{2024ApJ...962..190R} recently pointed out that the DBM is often quite ineffective in describing the proper propagation of CMEs in interplanetary space. In this study, a CME was observed by two radially aligned probes separated by a distance of just $0.13$ AU. Although the model predicted that the CME would decelerate, the velocity profiles measured by the two spacecraft instead revealed a residual acceleration, pointing to an additional force to the drag that overpowered its braking effect, and thus resulting in an increase in velocity. This work presents a more refined and realistic drag-based model, with the aim to overcome the limitations of current versions by introducing into the equation of motion describing the dynamic interaction of the CME with the solar wind an extra acceleration, representing any other forces involved. 

After obtaining and discussing the mathematical solutions of the resulting new equations of motion (\S\ref{sec:EDBM}.), the updated version is applied to the observation of the same CME already studied in \citet{2024ApJ...962..190R}, showing that it satisfactorily succeeds in describing its dynamic evolution, and thus becoming a significant breakthrough in the prediction of CME travel time in Space Weather studies (\S\ref{sec:numexp}.). Interpretation of which physical process(es) the additional acceleration is due to, is tentatively given in \S\ref{sec:concl}., where our conclusions are also offered. Computational details to derive formulae in \S\ref{sec:EDBM}. are summarized in the Appendix.

\section{The Extended Drag-Based Model} 
\label{sec:EDBM}
Let us consider a generalization of the DBM where the total net acceleration acting on the CME in the interplanetary phase is made of two contributions:
\begin{equation}
    \label{eqn:totacc}
    \ddot{r}=a_{\text{drag}}(r,t)+a_{\text{extra}}(r)\;,
\end{equation}
where $a_{\text{drag}}= -\gamma r^{-\alpha}|\dot{r} - w(r,t)|^k(\dot{r}-w(r,t))$, $k\in 2\mathbb{N}+1$, and $a_{\text{extra}}=ar^{-\beta}$, for appropriate exponents $\alpha,\beta>0$ and coefficients $\gamma>0$, $a\neq0$; $r=r(t)$ and $v(t)=\dot{r}(t)$ are the CME’s instantaneous radial position and speed (typically the CME front distance and front speed); $w(r,t)$ is the background solar wind speed given as a known function of position and time. Physically, the model describes the same
dynamics of the DBM perturbed by an extra (e.g., magneto-gravitational) force acting on the CME along the motion, altogether exponentially damped over distance.

In general, equation (\ref{eqn:totacc}) does not admit an analytical solution, which hampers the computation of $a_{\text{extra}}$ from time and space measurements, i.e., by solving a boundary value problem. A closed-form time solution of (\ref{eqn:totacc}) is possible by assuming $\alpha=\beta=0$ and a constant $w(r,t)\equiv w$, for any fixed odd integer $k$. Therefore, we introduce the Extended Drag-Based Model (EDBM hereafter) as the equation
\begin{equation}
\label{eqn:EDBM}
\ddot{r}=-\gamma|\dot{r}-w|(\dot{r}-w)+a\;,
\end{equation}
in which $k=1$ and which corresponds to a straightforward perturbation of the simplest form of the DBM studied in \cite{vrvsnak2013propagation}. The sign of $a\neq0$ in (\ref{eqn:EDBM}) establishes the form of the solution and the properties of the associated dynamical system. 

\begin{figure}
\centering
\includegraphics{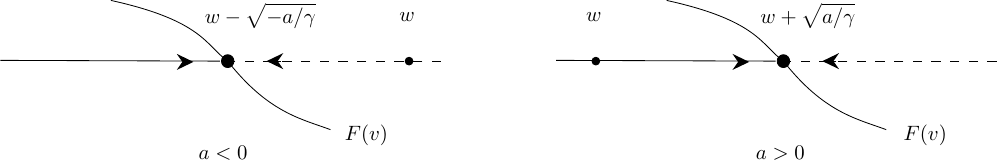}
\caption{Graphical representation of $F(v)\coloneqq-\gamma|v-w|(v-w)+a$ and local portrait of the speed dynamics around the stable equilibrium of the EDBM in the cases $a<0$ and $a>0$. The arrows define the positive sense of time for the evolution of $v(t)$. Solid and dashed lines correspond to $F>0$ (graph above the line, arrow pointing to the right) and $F<0$ (graph below the line, arrow pointing to the left), respectively.}
    \label{fig:EDBMdyn}
\end{figure}

We start from the equilibria, i.e.:
    \begin{itemize}
    \item if $a<0$, $v(t)\equiv w-\sqrt{-a/\gamma}$ is an asymptotically stable (in the future) constant solution of (\ref{eqn:EDBM}); thus, for $v_0>w-\sqrt{-a/\gamma}$ ($v_0<w-\sqrt{-a/\gamma}$) the CME monotonically decelerates (accelerates) for positive times;
   \item if $a>0$, $v(t)\equiv w+\sqrt{a/\gamma}$ is an asymptotically stable (in the future) constant solution of (\ref{eqn:EDBM}); thus, for $v_0>w+\sqrt{a/\gamma}$ ($v_0<w+\sqrt{a/\gamma}$) the CME monotonically decelerates (accelerates) for positive times.
   \end{itemize}
These assertions clarify the role of the acceleration term $a\neq0$: it shifts the asymptotic solution from $v=w$ (standard DBM) to $v=w\pm\sqrt{\pm a/\gamma}$ (EDBM). In contrast to the case $a=0$, this means that for $a>0$ ($a<0$) initial speeds below (above) the wind speed can increase (decrease) up (down) to $w$ and beyond. A schematic of the dynamics around the equilibrium points is provided in Figure \ref{fig:EDBMdyn}.

Equation (\ref{eqn:EDBM}) can be integrated from $0$ to $t>0$ to obtain explicit formulae for $v(t)$ and $r(t)$, given the initial conditions $v(0)=v_0$, $r(0)=r_0$. Depending on the choice of $v_0$, the solutions may be (differentiably) piecewise-defined for positive or negative values of $a$ due to the presence of the absolute value term $|\dot{r}-w|$, and present obvious symmetries in the form. Specifically,
\begin{description}
    \item[Case $a>0$]\phantom{.}\\

    \begin{itemize}
    \item if $v_0\le w$, then
    \begin{equation}
        \label{eqn:vsolaposv0lew}
        v(t)=
        \begin{cases}
        \displaystyle w+\sqrt{\frac{a}{\gamma}}\tan\left(\sqrt{a\gamma}t-\sigma_+\right)\;, & \text{for}\quad \displaystyle0\le t\le\frac{1}{\sqrt{a\gamma}}\sigma_+\\
        \displaystyle w+\sqrt{\frac{a}{\gamma}}\frac{e^{2(\sqrt{a\gamma}t-\sigma_+)}-1}{e^{2(\sqrt{a\gamma}t-\sigma_+)}+1}\;, & \text{for}\quad \displaystyle t>\frac{1}{\sqrt{a\gamma}}\sigma_+
        \end{cases}\;,
    \end{equation}
     \begin{equation}
        \label{eqn:rsolaposv0lew}
        r(t)=
        \begin{cases}
        \displaystyle wt+r_0-\frac{1}{\gamma}\ln\left(S_+\cos\left(\sqrt{a\gamma}t-\sigma_+\right)\right)\;, & \text{for}\quad \displaystyle0\le t\le\frac{1}{\sqrt{a\gamma}}\sigma_+\\
        \displaystyle \left(w-\sqrt{\frac{a}{\gamma}}\right)t+r_0+\frac{1}{\gamma}\left(\ln\left(\frac{e^{2(\sqrt{a\gamma}t-\sigma_+)}+1}{2S_+}\right)+\sigma_+\right)\;, & \text{for}\quad \displaystyle t>\frac{1}{\sqrt{a\gamma}}\sigma_+
        \end{cases}\;,
    \end{equation}
    where $\sigma_+\coloneqq\arctan(\sqrt{\gamma/a}(w-v_0))$ and $S_+\coloneqq \sqrt{(a+\gamma(v_0-w)^2)/a}$;

 \item if $v_0>w$, then
     \begin{equation}
        \label{eqn:vsolaposv0gw}
        v(t)=
        w+\sqrt{\frac{a}{\gamma}}\frac{A_+e^{2\sqrt{a\gamma}t}+B_+}{A_+e^{2\sqrt{a\gamma}t}-B_+}\;,\quad \text{for}\quad  t\ge 0\;,
    \end{equation}
    \begin{equation}
        \label{eqn:rsolaposv0gw}
        r(t)=\left(w-\sqrt{\frac{a}{\gamma}}\right)t+r_0
            +\frac{1}{\gamma}\ln\left(\frac{A_+e^{2\sqrt{a\gamma}t}-B_+}{2\sqrt{a}}\right)\;,\quad \text{for}\quad  t\ge 0\;,
    \end{equation}
    where $A_+\coloneqq\sqrt{\gamma}(v_0-w)+\sqrt{a}$ and $B_+\coloneqq\sqrt{\gamma}(v_0-w)-\sqrt{a}$.
    \end{itemize}

\item[Case $a<0$]\phantom{.}\\

    \begin{itemize}
    \item if $v_0\le w$, then
    \begin{equation}
    \label{eqn:vsolanegv0lew}
        v(t)=w+\sqrt{-\frac{a}{\gamma}}\frac{A_-e^{-2\sqrt{-a\gamma}t}+B_-}{A_-e^{-2\sqrt{-a\gamma}t}-B_-}\;,\quad \text{for}\quad  t\ge0\;,
    \end{equation}
    \begin{equation}
    \label{eqn:rsolanegv0lew}
        r(t)=\left(w-\sqrt{-\frac{a}{\gamma}}\right)t+r_0-\frac{1}{\gamma}\ln\left(\frac{A_-e^{-2\sqrt{-a\gamma}t}-B_-}{2\sqrt{-a}}\right)\;,\quad \text{for}\quad  t\ge0\;,
    \end{equation}
    where $A_-\coloneqq\sqrt{\gamma}(v_0-w)+\sqrt{-a}$ and $B_-\coloneqq\sqrt{\gamma}(v_0-w)-\sqrt{-a}$;    
    \item if $v_0> w$, then
    \begin{equation}
        \label{eqn:vsolanegv0gw}
        v(t)=\begin{cases}
            \displaystyle w-\sqrt{-\frac{a}{\gamma}}\tan\left(\sqrt{-a\gamma}t-\sigma_-\right)\;, &  \text{for}\quad \displaystyle 0\le t\le \frac{1}{\sqrt{-a\gamma}}\sigma_-\\
            \displaystyle w+\sqrt{-\frac{a}{\gamma}}\frac{e^{-2\left(\sqrt{-a \gamma}t-\sigma_-\right)}-1}{e^{-2\left(\sqrt{-a \gamma}t-\sigma_-\right)}+1}\;, &  \text{for}\quad \displaystyle t>\frac{1}{\sqrt{-a\gamma}}\sigma_-
        \end{cases}\;,
    \end{equation}
    \begin{equation}
        \label{eqn:rsolanegv0gw}
        r(t)=\begin{cases}
            \displaystyle wt+r_0+\frac{1}{\gamma}\ln\left(S_-\cos\left(\sqrt{-a\gamma}t-\sigma_-\right)\right)\;, & \displaystyle  \text{for}\quad  0\le t\le \frac{1}{\sqrt{-a\gamma}}\sigma_-\\
            \displaystyle\left(w-\sqrt{-\frac{a}{\gamma}}\right)t+r_0-\frac{1}{\gamma}\left(\ln\left(\frac{e^{-2(\sqrt{-a\gamma}t-\sigma_-)}+1}{2S_-}\right)-\sigma_-\right)\;, & \displaystyle  \text{for}\quad  t>\frac{1}{\sqrt{-a\gamma}}\sigma_-
        \end{cases}\;,
    \end{equation}
    where $\sigma_-\coloneqq\arctan(\sqrt{-\gamma/a}(v_0-w))$ and $S_-\coloneqq \sqrt{(a-\gamma(v_0-w)^2)/a}$.   
    \end{itemize}
    
    \end{description}
The expressions for the CME's speed reflect the dynamical behavior of Figure \ref{fig:EDBMdyn}: in (\ref{eqn:vsolaposv0lew}), $v(t)\le w$ in the former expression while $v(t)>w$ in the latter, and $v\to w+\sqrt{a/\gamma}$ as $t\to+\infty$; in (\ref{eqn:vsolaposv0gw}), $v(t)>w$ with $v\to w+\sqrt{a/\gamma}$ as $t\to+\infty$, and $v(t)$ is never smaller than or equal to $w$ for positive times; in (\ref{eqn:vsolanegv0lew}), $v(t)\le w$ with $v\to w-\sqrt{-a/\gamma}$ as $t\to+\infty$, and $v(t)$ is never larger than or equal to $w$ for positive times; finally, in (\ref{eqn:vsolanegv0gw}), $v(t)\ge w$ in the former expression while $v(t)<w$ in the latter, and $v\to w-\sqrt{-a/\gamma}$ as $t\to+\infty$.

The derivation of (\ref{eqn:vsolaposv0lew})--(\ref{eqn:rsolanegv0gw}) requires standard calculus techniques, whose details are given in Appendix \ref{sec:solEDBM}.

\section{Validation of the EDBM: the November 3\textsuperscript{rd} -- 5\textsuperscript{th} 2021 event}
\label{sec:numexp}
\begin{figure}
    \centering
    \includegraphics[width=0.495\linewidth]{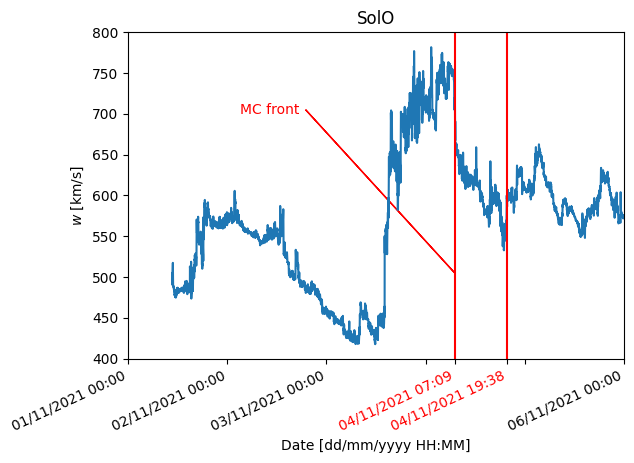}
    \includegraphics[width=0.495\linewidth]{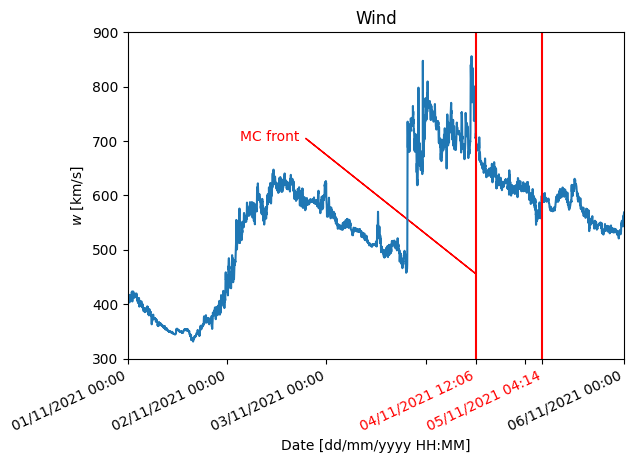}\\
    \includegraphics[width=0.495\linewidth]{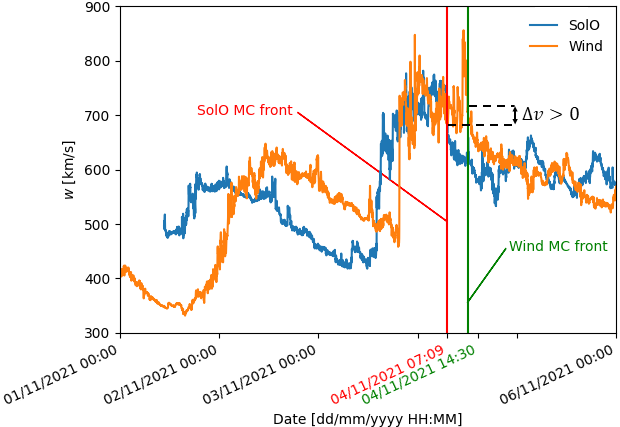}
    \caption{Time series of the ambient solar wind from November 1\textsuperscript{st} to November 6\textsuperscript{th} (UT) measured by the SolO Solar Wind Analyser (SWA) (left panel), by the Wind spacecraft (right panel), and both together (bottom panel). MC front limits are traced following \cite{regnault2024discrepancies} and $\Delta v$ is defined as in (\ref{eqn:amean}).} 
    \label{fig:windseries}
\end{figure}

As discussed in \textcolor{blue}{\S\ref{sec:intro}}., the closely spaced SolO-Wind detections of a CME of early November 2021 provided a reliable test-bed to assess the effectiveness of the EDBM. Indeed, in agreement with the analysis of \cite{regnault2024discrepancies}, the wind speed profiles measured by SolO and Wind (Figure \ref{fig:windseries}) suggested an acceleration of the CME Magnetic Cloud (MC) front from SolO to Wind (bottom panel) rather than the expected deceleration due to the drag force induced by the background solar wind. Although the physical reasons for this local behavior remain unclear (plausible interpretations are discussed in \textcolor{blue}{\S\ref{sec:concl}.}), in the following we applied the extended model described in \textcolor{blue}{\S\ref{sec:EDBM}.} against data collected at SolO and Wind locations $r_{\text{SolO}}$ and $r_{\text{Wind}}$ to estimate the additive acceleration $a>0$ between the two instruments. Specifically, given the measured MC front speeds $v_{\text{SolO}},v_{\text{Wind}}$ at times $t_{\text{SolO}},t_{\text{Wind}}$, respectively, we focused on the difference between the mean acceleration 
\begin{equation}
\label{eqn:amean}
a_{\text{mean}}=\frac{\Delta v}{\Delta t}=\frac{v_{\text{Wind}}-v_{\text{SolO}}}{t_{\text{Wind}}-t_{\text{SolO}}} \ \ ,
\end{equation}
which is an indicator of the approximate total measured acceleration exerted on the CME between the two spacecraft, and the model-dependent acceleration contributions $a_{\text{drag}}(\text{SolO})+a$ and $a_{\text{drag}}(\text{Wind})+a$, where
\begin{gather}
    \label{eqn:adragSolOWind}
    a_{\text{drag}}(\text{SolO})=-\gamma|v_{\text{SolO}}-w|(v_{\text{SolO}}-w)\;,\\
    a_{\text{drag}}(\text{Wind})=-\gamma|v_{\text{Wind}}-w|(v_{\text{Wind}}-w)\;.
\end{gather}

The main task, therefore, was to determine, from the solutions $v(t),r(t)$ in \textcolor{blue}{\S\ref{sec:EDBM}.}, the values of the extra-acceleration term $a$ that are compatible with the set of boundary values
\[
(v_{\text{SolO}},v_{\text{Wind}},r_{\text{SolO}},r_{\text{Wind}})=(690.86\text{ km/s},705.97\text{ km/s},0.85\text{ AU},0.98\text{ AU}) \ ,
\]
obtained from the data time series at initial time $t_{\text{SolO}}=0$ and final time $t_{\text{Wind}}=17820$ s, and the parameters $(\gamma,w)$. More specifically, we considered several experiments by choosing $w\in[400,800]$ km/s with incremental step $\Delta w=50$ km/s, and we used the same value $\gamma=0.24\times10^{-7}$ km$^{-1}$ as in \cite{regnault2024discrepancies}, compatible with the CME erupted on November 2\textsuperscript{nd} 2021 at 02:48 UT detected by the Solar and Heliospheric Observatory (SOHO) LASCO C2 coronograph (see, e.g., \cite{li2022solar}). Through a standard root-finding Newton-Raphson method \citep{suli2003introduction}, for every $w$ one has to search for the solution(s) (if any) of
\begin{equation}
\label{eqn:fv}
f_v(a)\coloneqq v(a;t_{\text{SolO}},t_{\text{Wind}},v_{\text{SolO}},w,\gamma)-v_{\text{Wind}}=0
\end{equation}
or
\begin{equation}
\label{eqn:fr}
f_r(a)\coloneqq r(a;t_{\text{SolO}},t_{\text{Wind}},v_{\text{SolO}},r_{\text{SolO}},w,\gamma)-r_{\text{Wind}}=0
\end{equation}
using formulae (\ref{eqn:vsolaposv0lew})--(\ref{eqn:rsolaposv0gw}) (case $a>0$). We initialized the root-finding algorithm with an initial guess of approximately the same order of magnitude of $|a_{\text{drag}}(\text{SolO})|,|a_{\text{drag}}(\text{Wind})|$, and iterate until convergence to a local positive value (did not the scheme converge, we would set $a=0$).

For the sake of simplicity, we applied this scheme to 
(\ref{eqn:fv}) and, since formula (\ref{eqn:vsolaposv0lew}) is case-defined and the time intervals depend on the unknown $a$, we eventually checked that for each experiment the corresponding time condition was fulfilled once $a$ is found (did not the time condition apply, we would reject the solution).

\begin{figure}
    \centering
    \includegraphics[width=0.494\linewidth]{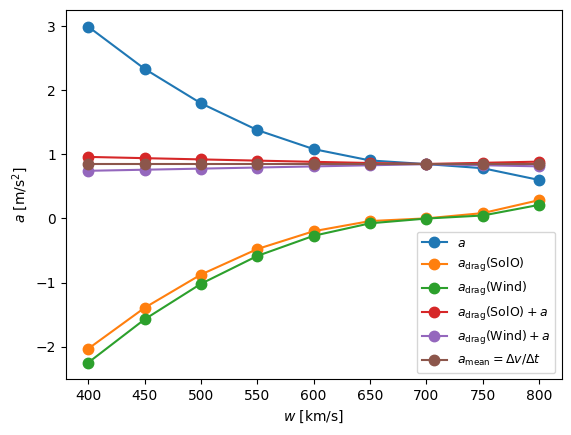}
    \includegraphics[width=0.499\linewidth]{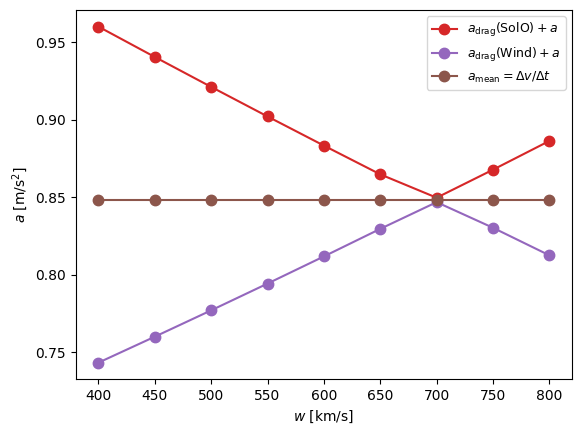}\\
    \hspace{2mm}\includegraphics[width=0.46\linewidth]{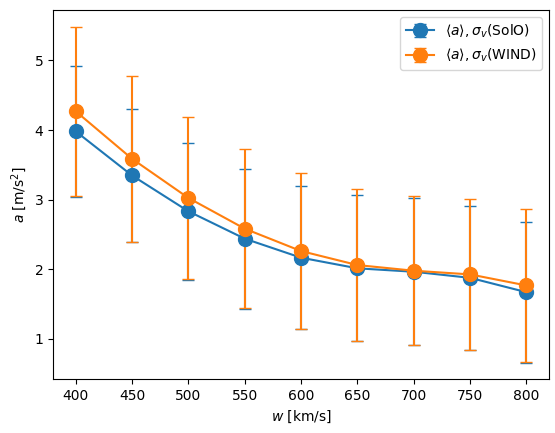}\hspace{2.5mm}
    \includegraphics[width=0.505\linewidth]{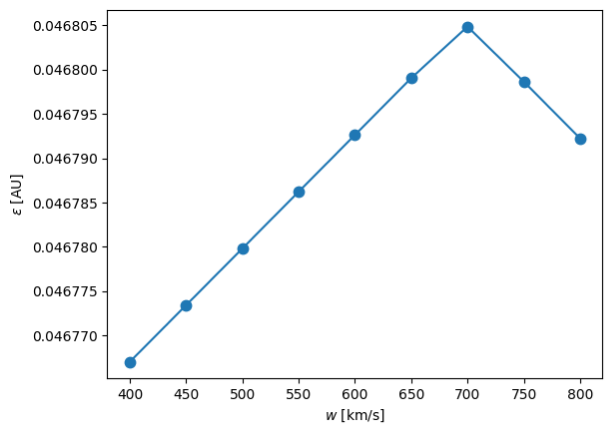}
    \caption{Modelling the November 3\textsuperscript{rd} -- 5\textsuperscript{th} 2021 event with the EDBM. Top left panel: the extra-acceleration term predicted by the EDBM (cyan line) is added to the acceleration terms predicted by the DBM at SolO (orange line) and Wind (green line) to obtain the red and purple lines that are compared to the experimental average acceleration from SolO to Wind (brown line). Top right panel: zoom on the experimental average and predicted accelerations. Bottom left panel: mean acceleration and corresponding standard deviation provided by the EDBM for 10 random realizations of the initial and final speeds (blue and orange lines, respectively). Bottom right panel: absolute error $\varepsilon=|f_r(a)|$ from (\ref{eqn:fr}) at Wind for the EDBM solutions $a$ obtained as in the top left panel using (\ref{eqn:fv}).}
    \label{fig:afromv}
\end{figure}

Figure \ref{fig:afromv} contains the results of this analysis. Specifically, in the top left panel a positive extra-acceleration $a$ was obtained for each choice of $w$ (cyan curve), as opposed to a drag deceleration at SolO and Wind until $w=700$ km/s (orange and green curves). The sum of the corresponding contributions provided two profiles that symmetrically fit the constant value $a_{\text{mean}}$ in (\ref{eqn:amean}) with a notable degree of accuracy, independently of the ambient solar wind (from the magnification in the top right panel, the maximum error committed is about $0.1$ m/s$^2$ attained at $w=400$ km/s). Furthemore, it is worth mentioning the optimality reached at $w=700$ km/s, with almost an exact match between the three accelerations: indeed, this is the value for the solar wind closest to $v_{\text{SolO}}$ and $v_{\text{Wind}}$.

The bottom panels of the same figure describe the outcomes of two further tests. First, we generated two sets containing ten values of the extra-acceleration $a$ computed for two sets of ten random realizations of the initial speed in the range $[v_{\text{SolO}} - 50,v_{\text{SolO}} + 10 ]$ km/s and of the final speed in the range $[v_{\text{Wind}}-10,v_{\text{Wind}}+50]$ km/s, respectively (the reason for this choice of ranges is two-fold: it guarantees that $v_{\text{SolO}}<v_{\text{Wind}}$, and a maximum error of $50$ km/s is plausible while accounting for the uncertainty on the temporal location of the MC boundary). The left panel contains average values $\langle a\rangle$ and the corresponding standard deviations $\sigma_v(\text{SolO}),\sigma_v(\text{Wind})$ computed over the two sets (these standard deviations stabilize after ten random realizations of the inital/final speeds). Note that $\sigma_v(\text{SolO})\approx\sigma_v(\text{Wind})\approx 1$ m/s$^2$ independently of $w$, and we coherently re-obtained the best agreement between the two profiles for $w=700$ km/s. Second, in the bottom right panel of Figure \ref{fig:afromv}, we computed the absolute error $\varepsilon=|f_r(a)|$ from (\ref{eqn:fr}) at Wind location for the solutions $a$ obtained as in the top left panel using (\ref{eqn:fv}). The overall error as a function of $w$ did not exceed $\varepsilon=0.046805$ AU (relative error $\approx 5\%$), attained at $w=700$ km/s. Note that for this value of the wind speed, we found, at the same time, the best outcome as far as $a$ is concerned, though the largest error on $r$. This suggests to rely on a trade-off strategy when fitting the real data either with the $v(t)$ model (equation (\ref{eqn:fv})) or the $r(t)$ model (equation (\ref{eqn:fr})).

In this respect, we infer that the EDBM can accurately describe the dynamics of a vast sample of interplanetary CMEs, especially of those excluded by the simplest form of the DBM, like the ones propelled beyond the solar wind speed (case with $w=700$ km/s in Figure \ref{fig:afromv}). Indeed, when $w=700$ km/s is assumed, the intermediate condition $v_{\text{SolO}}<w<v_{\text{Wind}}$ holds, which cannot be modelled using the classical DBM (see \S\ref{sec:EDBM}). 

\section{Discussion and conclusions}
\label{sec:concl}
As CMEs travel through interplanetary space, they can experience residual acceleration during their expansion due to several factors beyond solar wind drag. These include:

\begin{enumerate}
	\item \textbf{Magnetic forces}: CMEs are highly magnetized plasma structures, carrying their own magnetic field. As they expand into interplanetary space, their magnetic field interacts with the Sun’s Interplanetary Magnetic Field (IMF). This interaction can generate magnetic forces that can lead to residual acceleration, depending on the alignment and strength of the magnetic fields. The magnetic pressure from the Sun’s field, which decreases with distance, may provide a residual push on the CME as it expands.
	\item \textbf{Pressure gradients}: as CMEs move away from the Sun, they encounter regions of lower density and pressure. The difference between the internal pressure of the CME and the external pressure of the surrounding solar wind can cause the CME to continue expanding and accelerate. If the internal pressure of the CME remains higher than the external pressure for an extended period, this imbalance can drive residual acceleration during the CME’s expansion.
	\item \textbf{Internal magnetic reconfiguration}: the internal dynamics of a CME, including its magnetic tension forces and plasma flows, can also contribute to residual acceleration. Indeed, CMEs contain complex magnetic structures that can undergo reconfiguration or magnetic reconnection as they expand. These internal processes can release energy, contributing to the acceleration of the CME. For example, if magnetic loops within the CME reconnect, the release of magnetic energy could provide a push that accelerates the CME further into space.
	\item \textbf{Gravitational forces}: although relatively weak at large distances from the Sun, gravitational forces from the Sun can still play a role. Near the Sun, gravity decelerates the CME, but as it moves farther away, the influence of gravity decreases. If the CME has not yet reached a terminal velocity, the reduction in gravitational influence can result in a relative acceleration as the opposing force weakens.
	\item \textbf{Plasma and magnetic pressure balance}: the expansion of the CME involves the balance between plasma pressure and magnetic pressure within the CME and in the surrounding solar wind. As the CME expands and its internal pressure decreases, the balance between these pressures can change, leading to further acceleration. If the magnetic pressure within the CME remains relatively high, it could continue to push the CME outward.
\end{enumerate}

In general, these factors certainly contribute to the complex dynamics of CMEs as they travel through space, influencing their speed and trajectory beyond the initial influence of the solar wind. More specifically, the present study showed that the combination of these effects can be modelled by an extra-acceleration term that, when added to the drag force, contributes to explain the observations of a CME performed by SolO and Wind much more reliably than the standard DBM.

Understanding the processes itemized above is essential for predicting the behavior of CMEs and their potential impact on space weather and Earth’s environment. Disentangling which of these processes is currently at work in the evolution of CME under studio needs further analysis, which, as it is beyond the scope of the present work, is devoted to a future paper. Finally, it is also worth noting the interesting possibility of combining this extended drag-based model in the neural network developed by \citet{2023ApJ...954..151G}, so as to also refine their AI-based model and potentially make it even more predictive.

\section*{Acknowledgments}
SG was supported by the Programma Operativo Nazionale (PON) “Ricerca e Innovazione” 2014–2020. All authors acknowledge the support of the Fondazione Compagnia di San Paolo within the framework of the Artificial Intelligence Call for Proposals, AIxtreme project (ID Rol: 71708). AMM is also grateful to the HORIZON Europe ARCAFF Project, Grant No. 101082164. SG, MP and AMM are also grateful to the Gruppo Nazionale per il Calcolo Scientifico
- Istituto Nazionale di Alta Matematica (GNCS -
INdAM). MR is also grateful to the Gruppo Nazionale per la Fisica Matematica
- Istituto Nazionale di Alta Matematica (GNFM -
INdAM).

\section*{Appendix}

\subsection{Computation of the solutions of the EDBM}
\label{sec:solEDBM}

\subsubsection{$v_0,v\le w$}
\label{subsec:v0vlew}
Assume to integrate (\ref{eqn:EDBM}) in $[0,t]$ such that $v_0,v(t)\le w$:
\[
\int_{v_0}^v\frac{\text{d}v'}{a+\gamma(v'-w)^2}=\int_0^t\text{d}t'\;.
\]
Distinguishing between $a>0$ and $a<0$, we obtain two different primitives for the left-hand side and get:
\[
t=
\begin{cases}
    \displaystyle\frac{1}{\sqrt{a\gamma}}\left(\arctan\left(\sqrt{\frac{\gamma}{a}}(v-w)\right)-\arctan\left(\sqrt{\frac{\gamma}{a}}(v_0-w)\right)\right)\;,&a>0\\
    \displaystyle -\frac{1}{2\sqrt{-a\gamma}}\ln\left(\frac{(\sqrt{\gamma}(v-w)+\sqrt{-a})(\sqrt{\gamma}(v_0-w)-\sqrt{-a})}{(\sqrt{\gamma}(v_0-w)+\sqrt{-a})(\sqrt{\gamma}(v-w)-\sqrt{-a})}\right)\;,&a<0
\end{cases}\;;
\]
now solving for $v$ in both the expressions and setting $v(t)\le w$ yield formulae (\ref{eqn:vsolaposv0lew}) in the case $0\le t\le \arctan(\sqrt{\gamma/a}(w-v_0))/\sqrt{a\gamma}$ and (\ref{eqn:vsolanegv0lew}). 

As regards $r(t)$, for $a>0$ a second integration provides
\begin{align*}
r(t)&=\int_0^t v(t')\text{d}t'=wt+r_0+\sqrt{\frac{a}{\gamma}}\int_0^t\tan\left(\sqrt{a\gamma}t'+\arctan\left(\sqrt{\frac{\gamma}{a}}(v_0-w)\right)\right)\text{d}t'\\
&=wt+r_0-\frac{1}{\gamma}\ln\left(\left\lvert\cos\left(\sqrt{a\gamma}t+\arctan\left(\sqrt{\frac{\gamma}{a}}(v_0-w)\right)\right)\right\rvert\sqrt{\frac{a+\gamma(v_0-w)^2}{a}}\right)\;,
\end{align*}
using $|\cos x|=1/\sqrt{1+\tan^2x}$. In the interval $[0,\arctan(\sqrt{\gamma/a}(w-v_0))/\sqrt{a\gamma}]$ the cosine is positive, so we can remove the absolute value and obtain equation (\ref{eqn:rsolaposv0lew}) (first case).

For $a<0$, we conveniently set $A\coloneqq\sqrt{\gamma}(v_0-w)+\sqrt{-a}$, $B\coloneqq\sqrt{\gamma}(v_0-w)-\sqrt{-a}$ and $C\coloneqq-2\sqrt{-a\gamma}$. We have
\[
r(t)=\int_0^tv(t')\text{d}t'=wt+\int_0^t\frac{Ae^{Ct'}+B}{Ae^{Ct'}-B}\sqrt{-\frac{a}{\gamma}}\text{d}t'+r_0=\left(w-\sqrt{-\frac{a}{\gamma}}\right)t+r_0+\frac{2}{C}\sqrt{-\frac{a}{\gamma}}\ln\left\lvert\frac{Ae^{Ct}-B}{A-B}\right\rvert\;,
\]
where the integral is first computed by splitting the fraction as
\[
\frac{Ae^{Ct'}+B}{Ae^{Ct'}-B}=1+\frac{2B}{Ae^{Ct'}-B}\;,
\]
and then performing the two subsequent (monotonic) changes of variable $u=A\exp(Ct')-B$ and $U=1+B/u$. Again, we can disregard the absolute value for $t\ge0$, and replacing back the values of $A,B,C$ we get (\ref{eqn:rsolanegv0lew}).

\subsubsection{$v_0,v\ge w$}
\label{subsec:v0vgew}
The procedure is the same as in Appendix \ref{subsec:v0vlew}. Since now the integration of (\ref{eqn:EDBM}) reads
\[
\int_{v_0}^v\frac{\text{d}v'}{a-\gamma(v'-w)^2}=\int_0^t\text{d}t'\;,
\]
formulas derived for $v(t)$ and $r(t)$ are simply swapped for $a\lessgtr0$. Upon substituting $a\mapsto-a$, $\sqrt{a\gamma}\mapsto-\sqrt{-a\gamma}$ or the other way around depending on the sign of $a$, an analogous argument for time intervals, absolute values, and a corresponding re-definition of constants $A,B,C$ hold. This leads to equations (\ref{eqn:vsolaposv0gw}) ($a>0$), (\ref{eqn:vsolanegv0gw}) (first case, $a<0$) for $v(t)$, and (\ref{eqn:rsolaposv0gw}) ($a>0$), (\ref{eqn:rsolanegv0gw}) (first case, $a<0$) for $r(t)$.

\subsubsection{$v_0<w<v$}
\label{subsec:v0<w<v}
This time the integration of the EDBM gives rise to two contributions:
\[
\int_{v_0}^w\frac{\text{d}v'}{a+\gamma(v'-w)^2}+\int_w^v\frac{\text{d}v'}{a-\gamma(v'-w)^2}=t\;;
\]
in addition, from Figure \ref{fig:EDBMdyn}, the forward dynamics rules out the case $a<0$ (it is possible only backward in time) and requires $t>t_*\coloneqq\arctan(\sqrt{\gamma/a}(w-v_0))/\sqrt{a\gamma}$ to arrive at $v(t)>w$ (cf. Appendix \ref{subsec:v0vlew}). Then, we have
\[
\int_{w}^v\frac{\text{d}v'}{a-\gamma(v'-w)^2}=t-t_*\;,
\]
which is the case of Appendix \ref{subsec:v0vgew} with lower limit of integration equal to $w$. So, upon solving for $v$, we obtain expression (\ref{eqn:vsolaposv0lew}) in the case $t>t_*$.

Concerning $r(t)$, we need to integrate equation (\ref{eqn:vsolaposv0lew}) (second case) from $t_*$ to $t$:
\[
r(t)=w(t-t_*)+\int_{t_*}^t\frac{Ae^{Ct'}+B}{Ae^{Ct'}-B}\sqrt{\frac{a}{\gamma}}\text{d}t'+r_*\;,
\]
with $r_*=r(t_*)$, $A\coloneqq \exp(2\arctan(\sqrt{\gamma/a}(v_0-w)))$, $B\coloneqq -1$, $C\coloneqq2\sqrt{a\gamma}$. This relationship is formally identical to the one of $r(t)$ in Appendix \ref{subsec:v0vlew}, case $a<0$. We find
\[
r(t)=\left(w-\sqrt{\frac{a}{\gamma}}\right)(t-t_*)+r_*+\frac{1}{\gamma}\ln\left(\frac{Ae^{Ct}-B}{Ae^{Ct_*}-B}\right)\;.
\]
Lastly, we determine $r_*$ by enforcing continuity at $t=t_*$ with the former expression in (\ref{eqn:rsolaposv0lew}):
\[
r_*=wt_*+r_0-\frac{1}{\gamma}\ln\sqrt{\frac{a+\gamma(v_0-w)^2}{a}}\;;
\]
hence, replacing back in the expression of $r(t)$, we retrieve the latter of (\ref{eqn:rsolaposv0lew}).

\subsubsection{$v<w<v_0$}
\label{subsec:v<w<v0}

Following Appendix \ref{subsec:v0<w<v}, an analogous reasoning is applied in the case $a<0$. The corresponding sign adaptation of quantities involving $a$ (see Appendix \ref{subsec:v0vgew}) and the continuity requirement with the first relationship of equation (\ref{eqn:rsolanegv0gw}) produce formulae (\ref{eqn:vsolanegv0gw}), (\ref{eqn:rsolanegv0gw}) for $t>\arctan(\sqrt{-\gamma/a}(v_0-w))/\sqrt{-a\gamma}$. 

\bibliography{EDBMalignment}{}

\begin{thebibliography}{}
\expandafter\ifx\csname natexlab\endcsname\relax\def\natexlab#1{#1}\fi
\providecommand{\url}[1]{\href{#1}{#1}}
\providecommand{\dodoi}[1]{doi:~\href{http://doi.org/#1}{\nolinkurl{#1}}}
\providecommand{\doeprint}[1]{\href{http://ascl.net/#1}{\nolinkurl{http://ascl.net/#1}}}
\providecommand{\doarXiv}[1]{\href{https://arxiv.org/abs/#1}{\nolinkurl{https://arxiv.org/abs/#1}}}

\bibitem[{{Camporeale}(2019)}]{2019SpWea..17.1166C}
{Camporeale}, E. 2019, Space Weather, 17, 1166, \dodoi{10.1029/2018SW002061}

\bibitem[{{Chen}(2011)}]{2011LRSP....8....1C}
{Chen}, P.~F. 2011, Living Reviews in Solar Physics, 8, 1,
  \dodoi{10.12942/lrsp-2011-1}

\bibitem[{{Dumbovi{\'c}} {et~al.}(2018){Dumbovi{\'c}}, {{\v{C}}alogovi{\'c}},
  {Vr{\v{s}}nak}, {Temmer}, {Mays}, {Veronig}, \&
  {Piantschitsch}}]{2018ApJ...854..180D}
{Dumbovi{\'c}}, M., {{\v{C}}alogovi{\'c}}, J., {Vr{\v{s}}nak}, B., {et~al.}
  2018, The Astrophysical Journal, 854, 180, \dodoi{10.3847/1538-4357/aaaa66}

\bibitem[{{Gopalswamy}(2016)}]{2016GSL.....3....8G}
{Gopalswamy}, N. 2016, Geoscience Letters, 3, 8,
  \dodoi{10.1186/s40562-016-0039-2}

\bibitem[{{Gou} {et~al.}(2019){Gou}, {Liu}, {Kliem}, {Wang}, \&
  {Veronig}}]{2019SciA....5.7004G}
{Gou}, T., {Liu}, R., {Kliem}, B., {Wang}, Y., \& {Veronig}, A.~M. 2019,
  Science Advances, 5, 7004, \dodoi{10.1126/sciadv.aau7004}

\bibitem[{{Guastavino} {et~al.}(2023){Guastavino}, {Candiani}, {Bemporad},
  {Marchetti}, {Benvenuto}, {Massone}, {Mancuso}, {Susino}, {Telloni},
  {Fineschi}, \& {Piana}}]{2023ApJ...954..151G}
{Guastavino}, S., {Candiani}, V., {Bemporad}, A., {et~al.} 2023, The
  Astrophysical Journal, 954, 151, \dodoi{10.3847/1538-4357/ace62d}

\bibitem[{{Guastavino} {et~al.}(2024){Guastavino}, {Bahamazava},
  {Perracchione}, {Camattari}, {Audone}, {Telloni}, {Susino}, {Nicolini},
  {Fineschi}, {Piana}, \& {Massone}}]{2024ApJ...971...94G}
{Guastavino}, S., {Bahamazava}, K., {Perracchione}, E., {et~al.} 2024, The
  Astrophysical Journal, 971, 94, \dodoi{10.3847/1538-4357/ad5b57}

\bibitem[{Howard(2011)}]{howard2011coronal}
Howard, T. 2011, Coronal mass ejections: An introduction, Vol. 376 (Springer
  Science \& Business Media)

\bibitem[{{Jin} {et~al.}(2016){Jin}, {Schrijver}, {Cheung}, {DeRosa}, {Nitta},
  \& {Title}}]{2016ApJ...820...16J}
{Jin}, M., {Schrijver}, C.~J., {Cheung}, M.~C.~M., {et~al.} 2016, The
  Astrophysical Journal, 820, 16, \dodoi{10.3847/0004-637X/820/1/16}

\bibitem[{Li {et~al.}(2022)Li, Wang, Guo, \& Lyu}]{li2022solar}
Li, X., Wang, Y., Guo, J., \& Lyu, S. 2022, The Astrophysical Journal Letters,
  928, L6

\bibitem[{{Manchester} {et~al.}(2017){Manchester}, {Kilpua}, {Liu}, {Lugaz},
  {Riley}, {T{\"o}r{\"o}k}, \& {Vr{\v{s}}nak}}]{2017SSRv..212.1159M}
{Manchester}, W., {Kilpua}, E. K.~J., {Liu}, Y.~D., {et~al.} 2017, SSRv, 212,
  1159, \dodoi{10.1007/s11214-017-0394-0}

\bibitem[{{M{\"o}stl} {et~al.}(2018){M{\"o}stl}, {Amerstorfer}, {Palmerio},
  {Isavnin}, {Farrugia}, {Lowder}, {Winslow}, {Donnerer}, {Kilpua}, \&
  {Boakes}}]{2018SpWea..16..216M}
{M{\"o}stl}, C., {Amerstorfer}, T., {Palmerio}, E., {et~al.} 2018, SpWea, 16,
  216, \dodoi{10.1002/2017SW001735}

\bibitem[{{Napoletano} {et~al.}(2018){Napoletano}, {Forte}, {Del Moro},
  {Pietropaolo}, {Giovannelli}, \& {Berrilli}}]{2018JSWSC...8A..11N}
{Napoletano}, G., {Forte}, R., {Del Moro}, D., {et~al.} 2018, Journal of Space
  Weather and Space Climate, 8, A11, \dodoi{10.1051/swsc/2018003}

\bibitem[{Regnault {et~al.}(2024)Regnault, Al-Haddad, Lugaz, Farrugia, Yu,
  Zhuang, \& Davies}]{regnault2024discrepancies}
Regnault, F., Al-Haddad, N., Lugaz, N., {et~al.} 2024, The Astrophysical
  Journal, 962, 190

\bibitem[{{Regnault} {et~al.}(2024){Regnault}, {Al-Haddad}, {Lugaz},
  {Farrugia}, {Yu}, {Zhuang}, \& {Davies}}]{2024ApJ...962..190R}
{Regnault}, F., {Al-Haddad}, N., {Lugaz}, N., {et~al.} 2024, The Astrophysical
  Journal, 962, 190, \dodoi{10.3847/1538-4357/ad1883}

\bibitem[{S{\"u}li \& Mayers(2003)}]{suli2003introduction}
S{\"u}li, E., \& Mayers, D.~F. 2003, An introduction to numerical analysis
  (Cambridge university press)

\bibitem[{{Telloni} {et~al.}(2023){Telloni}, {Schiavo}, {Magli}, {Fineschi},
  {Guastavino}, {Nicolini}, {Susino}, {Giordano}, {Amadori}, {Candiani},
  {Massone}, \& {Piana}}]{2023ApJ...952..111T}
{Telloni}, D., {Schiavo}, M.~L., {Magli}, E., {et~al.} 2023, The Astrophysical
  Journal, 952, 111, \dodoi{10.3847/1538-4357/acdeea}

\bibitem[{Vr{\v{s}}nak {et~al.}(2013)Vr{\v{s}}nak, {\v{Z}}ic, Vrbanec, Temmer,
  Rollett, M{\"o}stl, Veronig, {\v{C}}alogovi{\'c}, Dumbovi{\'c}, Luli{\'c},
  {et~al.}}]{vrvsnak2013propagation}
Vr{\v{s}}nak, B., {\v{Z}}ic, T., Vrbanec, D., {et~al.} 2013, Solar physics,
  285, 295

\bibitem[{{Vr{\v{s}}nak} \& {Gopalswamy}(2002)}]{2002JGRA..107.1019V}
{Vr{\v{s}}nak}, B., \& {Gopalswamy}, N. 2002, JGRA, 107, 1019,
  \dodoi{10.1029/2001JA000120}

\bibitem[{{Vr{\v{s}}nak} {et~al.}(2013){Vr{\v{s}}nak}, {{\v{Z}}ic}, {Vrbanec},
  {Temmer}, {Rollett}, {M{\"o}stl}, {Veronig}, {{\v{C}}alogovi{\'c}},
  {Dumbovi{\'c}}, {Luli{\'c}}, {Moon}, \& {Shanmugaraju}}]{2013SoPh..285..295V}
{Vr{\v{s}}nak}, B., {{\v{Z}}ic}, T., {Vrbanec}, D., {et~al.} 2013, Solar
  Physics, 285, 295, \dodoi{10.1007/s11207-012-0035-4}

\bibitem[{{Webb} \& {Howard}(2012)}]{2012LRSP....9....3W}
{Webb}, D.~F., \& {Howard}, T.~A. 2012, LRSP, 9, 3,
  \dodoi{10.12942/lrsp-2012-3}

\end{thebibliography}
\bibliographystyle{aasjournal}

\end{document}